\documentclass[12pt]{article}
\usepackage{amssymb}
\usepackage{amsmath}

\newcommand{\bea}{\begin{eqnarray}}
\newcommand{\eea}{\end{eqnarray}}

\def\fr{\frac}

\def\bth{\bar{\theta}}

\def\dal{\dot{\alpha}}
\def\dbe{\dot{\beta}}
\def\dga{\dot{\gamma}}
\def\dde{\dot{\delta}}
\def\deps{\dot{\epsilon}}

\def\N{\mathcal{N}}
\def\A{\mathcal{A}}

\def\O{\mathcal{O}}

\def\M{\mathcal{M}}

\def\ovnab{\overline{\nabla}}

\def\2{\frac{1}{2}}

\def\spsi{\psi \hspace{-.50em}/}

\def\slpar{\partial \hspace{-.50em}/}

\begin{document}
\begin{titlepage}
\begin{flushright}
\begin{tabular}{l}
UTHEP-543\\
April  2007
\end{tabular}
\end{flushright}

\vspace{5mm}

\begin{center}
{\Large \bf  Superfield formulation of
 $4D$, $\mathcal{N}=1$ massless higher spin
 gauge field theory
and supermatrix model}
\baselineskip=24pt

\vspace{20mm}
{\large
Takashi Saitou\footnote{e-mail: saitout@het.ph.tsukuba.ac.jp}}\\
{\it Institute of Physics, University of Tsukuba,\\
     Tsukuba, Ibaraki 305-8571, Japan}\\
\vspace{20mm}
\end{center}

\begin{abstract}
We study the relation between
a supermatrix model and the free $4D$, $\mathcal{N}=1$ supersymmetric field
theory of a massless supermultiplet with
 spins (3, 5/2).
In order to do this,
we construct a superfield formulation of the theory.
We show that solutions of the equations of motion for the
supermultiplet (3, 5/2)
satisfy the equations of motion of a supermatrix model.

\end{abstract}
\end{titlepage}
\newpage


\section{Introduction}

It has been known that there are problems in the construction of
consistent interactions for
massless higher spin gauge fields,
though there are physically acceptable free field Lagrangian for them
\cite{Fronsdal:1978rb}
\cite{Fang:1978wz}
\cite{deWit:1979pe}
\cite{Vasiliev:1980as}.
The Lagrangian can be obtained by postulating
the gauge invariance which eliminates
unphysical degrees of freedom
\cite{Curtright:1979uz}.
A large amount of work has been done to construct interacting
massless higher spin gauge field theories.
Many attempts of them have
encountered difficulties associated with requirements of gauge
invariances
\cite{aragonedeser}
\cite{Berends:1979kg}
\cite{Aragone:1981yn}
\cite{Bengtsson:1986bz}
\cite{BBV},
though there are some consistent interacting
theories
\cite{Bekaert:2005vh}
\cite{Sezgin:2002rt}
\cite{Boulanger:2006gr}.
It is worthwhile to try to build  interacting
massless higher spin gauge field theories
by a new approach.

In the previous paper
\cite{Saitou:2006ca}, we have
studied  a matrix model as a new approach to formulate
massless higher spin gauge field theory.
As a first step towards constructing the
theory, we have shown that  the free equations of motion of bosonic
massless higher spin gauge fields can
      be derived from
those of the matrix model.
This is based on a new interpretation of matrix models
\cite{Hanada:2005vr}.
In \cite{Hanada:2005vr}, the authors have
introduced  a new interpretation of matrix models,
in which matrices represent differential operators on a curved
spacetime,
and have shown that the vacuum Einstein equation can be obtained
from
the equations of motion of a matrix model.
An advantage of this formalism is that
the matrix model possesses   gauge invariances manifestly, which are
embedded in
the unitary symmetry of the matrix model.
Therefore it is interesting to analyze interacting massless higher spin
gauge field theory using the matrix model.

In this paper, we study the relation between a supermatrix model
and
the free $4D$, $\N=1$
supersymmetric field theory of  massless supermultiplet with spins (3, 
5/2)
on the basis of \cite{Hanada:2006gg}.
In \cite{Hanada:2006gg},
the authors have extended the formalism in \cite{Hanada:2005vr}
to include supersymmetric field theories by
replacing matrices by
supermatrices in the matrix model.
Furthermore, they have shown that
solutions of the equations of motion for  the $4D$, $\N=1$ supergravity
satisfy the equations of motion of the supermatix model.
In this paper, we generalize their analysis 
  to higher spin gauge field theory of
a massless supermultiplet with spins (3, 5/2).
In order to do this,
 we construct
 both on-shell and  off-shell formulation for
the free $4D$, $\N=1$ supersymmetric field theory of
a massless   supermultiplet with spins (3, 5/2)
in terms of superfields.\footnote{
In four dimensional spacetime, all higher spin fields can be described
either by totally symmetric tensor or totally symmetric tensor-spinor
fields. In this paper, we will restrict our consideration
to  four dimensional field theories.}
We show that solutions of the equations of motion for the supermultiplet 
satisfy the equations of motion of the supermatrix model.
The formulations are quite similar to  the superfield formulations of
supergravity :
In  on-shell formulation, the equations of motion for
the  supermultiplet   can be
expressed as a constraint
on field strengths.
The superspace Bianchi identities subject to off-shell constraints
are solved and superfield strengths are expressed
by a set of superfields.

There are two ways to construct  superfield formulations of supersymmetric
field theories
\cite{Gates:1983nr} :
(1) One way is to study the off-shell representation to determine
the linearized formulation in terms of constraint-free superfields and
then construct covariant derivatives.
(2) Another way is to start by postulating the existence of covariant 
derivative,
and then
determine what constraints they must satisfy and solve them in terms of
a set of superfields.
An off-shell superfield formulation of
massless higher spin gauge field theory has been constructed
\cite{Kuzenko:1993jp}
in the way  (1).
The formulation we construct in this paper is in the way  (2).

There is another motivation for our study.
Massless higher spin fields
are expected to appear in the tensionless limit of
string theory, since mass squared of them are all
proportional to the string tension.
On the other hand, matrix models are expected to be a nonperturbative
formulation of string theory.
Therefore
our study
may
lead to further understanding
of nonperturbative aspects of string theory.

The organization of this paper is as follows.
In section 2, we briefly review the results of
\cite{Hanada:2006gg}.
In section 3, we construct an on-shell formulation of
a massless supermultiplet with spins (3, 5/2) in terms of superfield.
In section 4,
we study the relation between a supermatrix model and
the superfield formulation of the supermultiplet.
We show that 
solutions of the equations of motion for the supermultiplet 
satisfy the equations of motion of the supermatrix model.
Section 5 is devoted to conclusions and future works.
In appendix A, we summarize the on-shell constraints.
In appendix B, we give the explicit forms of
the superspace Bianchi identities subject to the on-shell constraints.
In appendix C, we give the results of
the off-shell superfield formulation.


\section{Supermatrix model}


\subsection{New interpretation of supermatrix model}

In \cite{Hanada:2005vr},
a new interpretation of matrix models has been proposed
in which matrices represent differential operators on
a  $D$ dimensional curved  space.
Matrices act as $Endomorphisms$ on a vector space,
which
means matrices map a vector space to itself.
On the contrary,
covariant derivatives map a tensor field of rank-$n$ to a tensor field
of rank-$(n+1)$.
In order to interpret differential operators as matrices,
we should prepare a vector space $V$ which contain at least tensor fields of
any rank.
In \cite{Hanada:2005vr},
the authors have shown that such a
space can be  given by the space of functions on the principal $Spin(D)$
bundle over a base manifold $M$.
Furthermore, they have considered the large $N$ reduced model 
of pure Yang-Mills theory as the matrix model.
Applying this new interpretation to the matrix model,
they have shown that the vacuum Einstein equation can be derived from
the equations of motion of the matrix model.

However, supergravity cannot be embedded in the usual bosonic matrix model
because there are no Grassmann variables in the matrix model.
Thus, in order to describe supergravity by  matrix models,
we need to extend $V$ to include Grassmann variables.
It has been shown that  this is implemented by
extending manifold $M$ to a supermanifold $\M$
 and
taking $V$ to be the space of functions on the principal $Spin(D)$ bundle
over $\M$
\cite{Hanada:2006gg}.
In this extension, matrices are replaced by supermatrices, and
covariant derivatives are replaced by supercovariant derivatives.
In \cite{Hanada:2006gg}, the authors have considered the supermatrix model
which
is obtained by replacing matrices by
supermatrices in
the Large-$N$ reduced model of pure Yang-Mills action\footnote{
We can consider the supersymmetric version of the supermatrix model,
which is the supermatrix generalization of IIB matrix model
\cite{Ishibashi:1996xs},
\begin{eqnarray}
S=-\frac{1}{4}{\rm Str}{[}\A_a,\A_b{]}{[}\A^a,\A^b{]}
+\frac{1}{2}{\rm Str}\overline{\Psi}\gamma^a{[}\A_a,\Psi{]},
\label{eq:actionsupmat1}
\end{eqnarray}
where $\A_a$ are Grassmann even supermatrices and $\Psi$ are Grassmann
odd supermatrices.
This action has global $\N=2$ supersymmetry,
but we could not understand the meanings of global $\N=2$
supersymmetry
of this model in the new interpretation.
Thus, we restrict our consideration to (\ref{eq:superpym}).
}
,
\begin{eqnarray}
S=-\frac{1}{4}{\rm Str}\Big(
[\A_a,\A_b][\A^a,\A^b]\Big),
\label{eq:superpym}
\end{eqnarray}
where $\A_a$ are hermitian and Grassmann even supermatrices with vector
index.
This action has $SO(D)$ Lorentz symmetry and superunitary symmetry
$U(N_e|N_o)$.\footnote{
An  even supermatrix ${\cal A}$
can be written as
\begin{eqnarray}
\A=\left( \begin{array}{cc}
A_1 & B_1 \\
B_2 & A_2
\end{array}
\right),
\end{eqnarray}
where $A_1$ are $N_e \times N_e$, $A_2$ are $N_o \times N_o$
bosonic matrices and $B_1$ are $N_o \times N_e$, $B_2$ are $N_e \times
N_o$ fermionic matrices.}
Applying the new interpretation to
this supermatrix model, the authors have  shown that
solutions of the equations of motion for the $D=4$, $\N=1$ supergravity
satisfy the equations of motion of the supermatrix
model.


\subsection{Massless higher spin fields}

Let us see that there is a possibility that the supermatrix model involves
the degrees of freedom of
massless higher spin gauge fields.
Before we begin discussing massless higher spin fields, we
explain our notations.
The coordinates of a superspace $\M$ are  expressed as 
$z^M=(x^m,\theta^\mu)$,
 where $x^m (m=1,\cdots ,D)$ are bosonic and $\theta^\mu
 (\mu=1,\cdots , D_s)$ are fermionic components.
$D_s$ is the dimension of spinor representation of $Spin(D)$.
Letters $M=(m,\mu)$ denote curved space indices and
$A=(a,\alpha)$ denote local Lorentz indices.
The supercovariant derivative
$\nabla_A$
is defined as
\begin{eqnarray}
\nabla_A=e_A{}^M (z)(\partial_M+\omega_M{}^{bc}(z)\O_{bc}),
\end{eqnarray}
where $e_A{}^M(z)$ is the supervielbein and $\omega_M{}^{bc}(z)$
is the superspin connection.
Notice that $\nabla_A$ maps a rank-$n$ tensor to a rank-$(n+1)$ tensor and
$\O_{ab}$ acts on the local Lorentz indices of these tensors. Therefore
we have
\begin{eqnarray}
{[}\O_{ab},\nabla_c{]}&=&
\2(\delta_{ac}\nabla_b-\delta_{bc}\nabla_a),\\
{[}\O_{ab},\nabla_\alpha{]}&=& (\gamma_{ab})_\alpha{}^\beta \nabla_\beta,
\end{eqnarray}
in this setting, which will be used later.

Since each component of supermatrices $\A_a$
acts on the functions on the principal $Spin(D)$ bundle over $\M$ as an
$Endomorphism$, in general, $\A_a$ can be expanded as
 \begin{eqnarray}
 \A_a = i\nabla_a +a_a(z) +\frac{i}{2}\{b_a{}^B(z),\nabla_B\}
+\frac{i}{2}\{\omega_a{}^{bc}(z),\O_{bc}\}+
\frac{i^2}{2}\{e_a{}^{BC}(z),\nabla_B\nabla_C\}+\cdots,
\label{eq:expmat1}
\end{eqnarray}
where $i$ and anticommutator $\{\}$ are introduced to make $\A_a$
hermitian supermatrices. Terms higher than first order with respect to
the operators
$\nabla_A$ and $\O_{ab}$
 can be taken to be symmetric (or antisymmetric)
 under permutations of the operators,
because antisymmetric (or symmetric) part can be absorbed
in the term that is the lower order in $\nabla_A$ and $\O_{ab}$.
We consider the expansion as a sum of homogeneous polynomials
of $\nabla_A$ and $\O_{ab}$, whose coefficients are identified with
massless higher spin gauge fields.
Coefficients of the first order homogeneous polynomial will express
gauge fields of the supermultiplet $(2,3/2)$, and those of the
 second order one will express gauge fields of the
supermultiplet $(3, 5/2)$ and
so on.
The number of independent components of higher spin gauge fields
grows rapidly with degree in $\nabla_A$ and $\O_{ab}$.

If the supermatrix model has the degrees of
freedom of massless higher spin fields,
the gauge symmetries associated with those fields should be included.
We find that  the symmetries can be  realized as
the superunitary symmetry of the
supermatrix model.
Originally,
the superunitary symmetry is written as
\begin{eqnarray}
\delta \A_a&=&
i {[}\Lambda ,\A_a{]}, \label{eq:mmunsym}
\end{eqnarray}
where $\Lambda$ is a $N \times N$ hermitian supermatrix.
In the new interpretation,
$\Lambda$ becomes a scalar operator expanded in
terms of $\nabla_A$ and $\O_{ab}$.

Let us check
how gauge transformations are generated by $\Lambda$ in the case
of
the supermultiplet $(3,5/2)$.
In order to deal with this case, we need to keep track of
the following terms
\begin{eqnarray}
\A_a= i\nabla_a + \frac{(i)^2}{2} e_{a,}{}^{bc}(\nabla_b\nabla_c
+\nabla_c\nabla_b)+ \frac{(i)^2}{2} e_{a,}{}^{c\gamma}(\nabla_c\nabla_\gamma
+\nabla_\gamma\nabla_c)+\cdots.
\end{eqnarray}
We take $\Lambda$ as $\Lambda=\lambda^{c\gamma}(\nabla_c\nabla_\gamma
+\nabla_\gamma\nabla_c)$, then (\ref{eq:mmunsym}) becomes
\begin{eqnarray}
\delta \A_a=(\nabla_a \lambda^{c\gamma})(\nabla_c\nabla_\gamma
+\nabla_\gamma\nabla_c) +\cdots.
\end{eqnarray}
Thus $e_{a,}{}^{c\gamma}$ transforms as
\begin{eqnarray}
\delta e_{a,}{}^{c\gamma}
= \nabla_a \lambda^{c\gamma}+\cdots.
\end{eqnarray}
This can be considered as the  supergauge transformation for the spin-5/2 
field.


\subsection{Superfield formulation}

In order to study the relation between
the supermatrix model and supersymmetric field theories
of massless higher spin supermultiplets,
we should  compare the equations of  motion of
the supermatrix model with
those of  the supermultiplets.
Since the local fields which appear in (\ref{eq:expmat1}) live in 
superspace,
the equations of motion of the supermatrix model
are written in terms of superfields.
Thus, we should write the equations of motion of
massless higher spin supermultiplets in terms of superfields
to compare with the results of the supermatrix model.
Namely, we should construct a superfield formulation of 
the supermultiplets.
Recall that for supergravity, we can construct superfield formulation by
starting with the supercovariant derivative $\nabla_A$,
and then imposing constraints on the field strengths which are defined as
the coefficients of the operators $\nabla_A$ and $\O_{ab}$ in
the commutators of $\nabla_A$,
\begin{eqnarray}
{[}\nabla_A, \nabla_B\} = C_{AB}{}^{C}(z)\nabla_C + R_{AB}{}^{cd}(z)\O_{cd}.
\end{eqnarray}
The equations of motion for supergravity
 are expressed as a constraint on the torsion tensor.
It seems that
we can construct superfield formulation
for supermultiplets of massless higher spin fields
in the same way.
We consider (\ref{eq:expmat1}) as  the supercovariant derivative 
 with vector index.
We postulate the existence of  the supermatrices with spinor index,
 \begin{eqnarray}
 \A_\alpha = i\nabla_\alpha +a_\alpha(z) 
+\frac{i}{2}\{b_\alpha{}^B(z),\nabla_B\}
+\frac{i}{2}\{\omega_\alpha{}^{bc}(z),\O_{bc}\}+
\frac{i^2}{2}\{e_\alpha{}^{BC}(z),\nabla_B\nabla_C\}+\cdots.
\label{eq:expmat2}
\end{eqnarray}
We can  regard that
the supermatrices  (\ref{eq:expmat1})  and  (\ref{eq:expmat2})
as the supercovariant derivative for
 massless higher spin fields. 
The field strengths are defined as
the coefficients of the operators $\nabla_A$ and $\O_{ab}$ in
the commutators of 
(\ref{eq:expmat1})  and  (\ref{eq:expmat2}).

In the next section, we will construct superfield formulation
for the free theory of
a massless supermultiplet with spins (3, 5/2)
using this supercovariant derivative.
Then, we will compare the results with those of the supermatrix model.


\section{Superfield formalism of massless supermultiplet (3,5/2)}

Now let us construct a superfield formulation of
the free  $4D$, $\N=1$ supersymmetric field theory of
a massless supermultiplet with spins
$(3,5/2)$
by starting from the supercovariant derivative  (\ref{eq:expmat1})
and (\ref{eq:expmat2}).
The construction is similar to supergravity : 
we should impose constraints on superfields.
One difference is in
fixing gauge symmetries which act on auxiliary fields to
eliminate auxiliary component fields.
As in the case of
 supergravity we can construct  on-shell and off-shell formulations.
In this section we restrict attention to on-shell
formulation.
We give the results of  off-shell formulation in appendix C.

Before we begin our analysis,
we review some facts about
the component formalism of
massless higher spin gauge fields
\cite{Curtright:1979uz}.
Totally symmetric tensor field of rank-$s$ $\phi_{a_1\cdots a_s}(x)$
and tensor-spinor field of rank-$(s-1)$
$\psi_{a_1\cdots a_{s-1},\alpha}(x)$ are
used to express massless boson and fermion system of a supermultiplet
with
spins $(s, s-\frac{1}{2})$.\footnote{
From this section, Latin  letters run from 1 to $4$
and denote flat spacetime vector indices, and
Greek letters run from 1 to 4 and denote spinor indices.}
We can construct the theory of $\phi_{a_1\cdots a_s}(x)$ and
$\psi_{a_1\cdots a_{s-1}}(x)$ by requiring that the theory has the
proper gauge symmetries.
Let us postulate that the theory is invariant
under  the following gauge transformations :
\begin{eqnarray}
\delta  \phi_{a_1\cdots a_{s}}(x)&=&
\partial_{(a_1}\lambda_{a_2\cdots a_{s})}(x),
\label{eq:symtengaugetdb}\\
\delta \psi_{a_1\cdots a_{s-1},\alpha}(x)
&=&
\partial_{(a_1} \xi_{a_2\cdots a_{s-1}),\alpha}(x),
\label{eq:symtengaugetd}
\end{eqnarray}
where the bracket $()$ denotes symmetrization of the flat spacetime
indices.
The gauge parameters $\lambda_{a_1\cdots a_{s-1}}(x)$ and $\xi_{a_1\cdots
a_{s-2},
\alpha}(x)$ are
rank-$(s-1)$ totally symmetric tensor with the traceless condition
$\xi^b{}_{b a_1\cdots a_{s-3}}=0$ and
totally symmetric tensor-spinor with the gamma-traceless condition
$(\gamma^b)_{\alpha}{}^\beta
\xi_{b a_1\cdots a_{s-3},\beta}=0$, respectively.
If we impose the additional double traceless constraints $\phi''_{a_1\cdots
a_{s-5}}=0$ and
triple-gamma traceless constraints $\spsi'_{a_1\cdots a_{s-5}}=0$,
we can find that the gauge invariant free equations of
motion for $\phi_{a_1\cdots a_s}(x)$ and  $\psi_{a_1\cdots
a_{s-1},\alpha}(x)$ are
\begin{eqnarray}
W_{a_1\cdots a_s}
&\equiv&
\Box \phi_{a_1\cdots a_s}-s\partial_{(a_1} (\partial \cdot \phi)_{a_2\cdots
a_s)}
+s(s-1)\partial_{(a_1} \partial_{a_2} \phi'_{a_3\cdots a_s)}
=0, \label{eq:3symtensoreom}
\\
Q_{a_1\cdots a_{s-1},\alpha}
&\equiv&
(\slpar)_{\alpha\beta}\psi_{a_1\cdots a_{s-1},}{}^\beta
-(s-1)\partial_{(a_1} \spsi_{a_2\cdots a_{s-1}),\alpha}
=0,\label{eq:spin5/2eom}
\end{eqnarray}
where we use the notations
$(\partial\cdot\phi)_{a_1\cdots a_{s-1}}
=\partial_b\phi^b{}_{a_1\cdots a_{s-1}}$,
 $\phi'_{a_1\cdots a_{s-3}}=\phi^b{}_{ba_1\cdots a_{s-2}}$
,$(\gamma^a)_{\alpha\beta}\partial_a=(\slpar)_{\alpha\beta}$ and
 $\spsi_{a,\alpha}=(\gamma^b)_{\alpha\beta}\psi_{ba_1\cdots
a_{s-2}}{}^\beta$.

The conventional formulations for free totally symmetric tensor and
tensor-spinor gauge
fields have been  originally derived by Fronsdal
\cite{Fronsdal:1978rb} and Fang-Fronsdal \cite{Fang:1978wz},
respectively.


\subsection{On-shell formulation of supermultiplet $(3, \frac{5}{2})$}

In order to deal with the supermultiplet (3, 5/2)
in terms of superfield,
we keep track of the second order homogeneous polynomial
of the operators $\partial_a$, $\nabla_\alpha$ and $\O_{ab}$ in $\A_A$ :
\begin{eqnarray}
\A_A &=& i\nabla_A
+i^2 e_A{}^{CD}(z)\nabla_C\nabla_D
\nonumber\\
&&
\hspace{30pt}
+\frac{i^2}{2}\omega_A{}^{C,de}(z)(\nabla_C\O_{de}+\O_{de}\nabla_C)
\nonumber\\
&&
\hspace{35pt}
+\frac{i^2}{2}\Omega_A{}^{cd,ef}(z)(\O_{cd}\O_{ef}+\O_{ef}\O_{cd}),
\label{eq:smmat}
\end{eqnarray}
where the supercovariant derivative in flat superspace is defined as
 \begin{eqnarray}
 \nabla_a = \partial_a, \quad \nabla_\alpha=\frac{\partial}{\partial
 \theta^\alpha}+i(\gamma^a)_{\alpha\beta}
 \theta^\beta \partial_a.
 \end{eqnarray}
The commutation relations of the operators are given by
\begin{eqnarray}
&& [\partial_a,\partial_b]= 0 ,\qquad
 {[} \partial_a,\nabla_\alpha {]} = 0, \qquad  \{\nabla_\alpha,\nabla_\beta
\}=
2i(\gamma^a)_{\alpha\beta}\nabla_a,
\\ &&
{[}\O_{ab},\partial_c{]}= \frac{1}{2}(\delta_{ac}\partial_b
-\delta_{bc}\partial_a),\qquad
{[}\O_{ab},\nabla_\alpha{]}= (\gamma_{ab})_\alpha{}^\beta \nabla_\beta.
 \end{eqnarray}
In this and the next subsection,
in order to deal with free field theories we keep only terms linear
with respect to the component fields  and use the
flat supercovariant derivatives defined above.

The dynamical fields which describe the supermultiplet (3, 5/2) are
expressed as
\begin{eqnarray}
\phi_{abc}(x) &=& \frac{1}{3} \big( e_{a,bc}(z)
+e_{b,ca}(z) +e_{c,ab}(z) \big) \Big|_{\theta=0}, \label{eq:dyn3}\\
\psi_{ab,\alpha}(x) &=&
\frac{1}{2}\big( e_{a,b\alpha}(z)+e_{b,a\alpha}(z)
\big)\Big|_{\theta=0} ,\label{eq:dyn52}
\end{eqnarray}
where $e_{a,bc}(z) $ are  the coefficients of $\partial_b\partial_c$  and
$e_{a,b\alpha}(z)$ are the  coefficients of $\partial_b\nabla_\alpha$ in
(\ref{eq:smmat}).
As we will see, these relations can be understood
by looking at the gauge transformation properties of these fields.
Local superfields appearing in (\ref{eq:smmat})
have too many  unphysical degrees of freedom to describe
the physical system of the massless supermultiplet (3, 5/2).
Thus, in order to construct superfield formulation we must eliminate
all the unphysical degrees of freedom.
This is implemented by carrying out the following two procedures
 \begin{itemize}
\item Imposing constraints on the superfields.
\item Fixing  the gauge symmetries.
\end{itemize}
We will perform these procedures in order.


\subsubsection{Constraints}

There are three types of constraints.\footnote{The constraints we impose in 
this subsection
 are summarized in appendix A.
The superspace Bianchi
identities subject to the on-shell constraints are given in appendix B.}

\begin{itemize}
\item[{\bf 1.}] The first type of constraints are summarized as follows :
\begin{eqnarray}
&& e_{a,b}{}^b(z)=0, \label{eq:1cnrt1}\\
&& (\gamma^b)_\alpha{}^\beta e_{a,b\beta}(z)=0. \label{eq:1cnrt2}
\end{eqnarray}
As we will see later,
we find the gauge transformation laws for $e_{a,bc}$ :
$\delta e_{a,bc}=\partial_a \lambda_{bc}$, and for $e_{a,b\beta}$
: $\delta e_{a,b\beta}=\partial_a \xi_{b\beta}$.
Thus, these constraints are necessary to be consistent
with the traceless constraints on the gauge parameters
$\lambda_b{}^b=0$ and $(\gamma^b)_\alpha{}^\beta \xi_{b,\beta}=0$.

\item[{\bf 2.}] The second type of constraints are imposed on the field
strengths.
Field strengths are the coefficients of the operators in the
commutators of $\A_a$ and $\A_\alpha$, whose
general expressions are given by the following forms :
\begin{eqnarray}
{[}\A_A,\A_B{]}&=&
-iC_{AB}{}^{CD}(z)\nabla_C\nabla_D
\nonumber\\
&& -\frac{i}{2}R_{AB}{}^{C,de}(z)(\nabla_C\O_{de}+\O_{de}\nabla_C)
\nonumber\\
&&
-\frac{i}{2}F_{AB}{}^{cd,ef}(z)(\O_{cd}\O_{ef}+\O_{ef}\O_{cd}).
\end{eqnarray}
$C_{AB,CD}(z)$ are similar to  the torsion tensor in supergravities
because they include the first order
derivatives of the vielbein fields $e_{a,bc}$ and $e_{a,b\alpha}$
with respect to $x$.
$R_{AB,C,de}(z)$ are similar to  the curvature tensor because
they include the first order
derivatives of the connection $\omega_{A,B,cd}$
with respect to $x$.\footnote{
In analogy with   superfield formulations of supergravities,
we can regard that the fields $e_{A,BC}(z)$ and $\omega_{A,B,cd}(z)$
are higher spin generalization of supervielbein and superspin
connection.   }
$F_{AB,cd,ef}(z)$ have no analogy in  supergravities because
they appear only for spin larger than 2.

We choose the following constraints :
\begin{eqnarray}
&& C_{ab,}{}^{cd}=
C_{ab}{}^{\gamma\delta}=0,
\qquad
C_{a\alpha,}{}^{cd}=
C_{a\alpha,}{}^{\gamma\delta}= 0,\label{eq:c23m}\\
&&
C_{\alpha\beta,}{}^{cd}=2i(\gamma^a)_{\alpha\beta}e_a{}^{cd}\label{eq:c31m}
,\quad
C_{\alpha\beta,}{}^{c\gamma}=2i(\gamma^a)_{\alpha\beta}e_{a}{}^{c\gamma}
\label{eq:c32m},\quad
C_{\alpha\beta,}{}^{\gamma\delta}=2i(\gamma^a)_{\alpha\beta}
e_{a}{}^{\gamma\delta},\label{eq:c33m}
\end{eqnarray}
\begin{eqnarray}
R_{ab}{}^{\gamma,cd}= 0,\quad
R_{a\alpha}{}^{\gamma,cd}=0,\quad
R_{\alpha\beta}{}^{\gamma,cd}=2i(\gamma^a)_{\alpha\beta}
\omega_{a,}{}^{\gamma,cd},\label{eq:c34m}
\end{eqnarray}
\begin{eqnarray}
F_{ab}{}^{cd,ef}=0, \quad
F_{a\alpha}{}^{cd,ef}=0  ,\quad
F_{\alpha\beta}{}^{cd,ef}=2i(\gamma^a)_{\alpha\beta}\Omega_{a,}{}^{cd,ef}.
\label{eq:c35m}
\end{eqnarray}
The equations of motion for the supermultiplet can be expressed as the
constraints on the field strength :
\begin{eqnarray}
C_{a\alpha}{}^{c\gamma}&=&0. \label{eq:eomc1m}
\end{eqnarray}

\item[{\bf 3.}] The third type of constraints are imposed
for the equations of motion to be symmetric
under
permutation of the vector indices.
The constraint $F_{ab,cd,ef}=0$ implies that $\Omega_{a,bc,de}(z)$ can be
written as a pure gauge like configuration
\begin{eqnarray}
\Omega_{a,bc,de}(z)=\partial_a \chi_{bc,de}(z),
\end{eqnarray}
where the parameter $\chi_{bc,de}$ satisfies
$\chi_{bc,de}=-\chi_{cb,de}=-\chi_{bc,ed}$.
In order to make the equation of motion  for the spin-3 field
to be symmetric under
permutations of the vector indices, we should impose
\begin{eqnarray}
\chi_{bc,de}(z)=-\frac{1}{3}\omega_{[b,c],de}(z). \label{eq:3cnrt1}
\end{eqnarray}
The constraint $R_{ab,\gamma,de}=0$ implies that $\omega_{a,\gamma,cd}(z)$
can be written as a pure gauge configuration
\begin{eqnarray}
\omega_{a,\gamma,cd}(z)=\partial_a \eta_{cd,\gamma}(z),
\end{eqnarray}
where $\eta_{cd,\gamma}$ satisfies  $\eta_{cd,\gamma}=-\eta_{dc,\gamma}$.
In order to make the equation of motion for the spin-$\frac{5}{2}$ field
  to be symmetric
under permutations of the vector indices, we should impose
\begin{eqnarray}
\eta_{cd,\gamma}(z)=-e_{[c,d],\gamma}(z). \label{eq:3cnrt2}
\end{eqnarray}
\end{itemize}

\vspace{\baselineskip}

Imposing the constraints (\ref{eq:1cnrt1}), (\ref{eq:1cnrt2}),
(\ref{eq:c23m})-(\ref{eq:eomc1m}), (\ref{eq:3cnrt1}) and (\ref{eq:3cnrt2}),
we obtain
\begin{eqnarray}
{[} \A_a,\A_b
{]}
&=&-iC_{ab,}{}^{c\gamma}(z)\partial_{c}\nabla_\gamma-\frac{i}{2}
R_{ab,}{}^{c,de}(z)
(\partial_c\O_{de}+\O_{de}\partial_c) ,\\
{[}
\A_a,\A_\alpha {]} &=& -\frac{i}{2}R_{a\alpha}{}^{c,de}(z)
(\partial_c\O_{de}
+\O_{de}\partial_c) \nonumber\\
\{\A_\alpha,\A_\beta\}&=& -2i(\gamma^a)\A_a,
\end{eqnarray}
where we use the fact that $R_{\alpha\beta}{}^{\gamma,cd}$ can be
written as
$R_{\alpha\beta}{}^{\gamma,cd}=2i(\gamma^a)_{\alpha\beta}\omega_{a,}{}^{\gamma,cd}
+\tilde{R}_{\alpha\beta}{}^{\gamma,cd}$.
$\tilde{R}_{\alpha\beta}{}^{\gamma,cd}=0$ follows from
the superspace Bianchi identity (\ref{eq:4b31}).

With all these constraints, using the superspace
Bianchi identities we can show that the equations of motion for
spin-$\frac{5}{2}$ field
\begin{eqnarray}
(\gamma^a)_{\alpha\beta} C_{ab,c,}{}^{\beta}=0,\label{eq:spin5/2eomc}
\end{eqnarray}
and for spin-3 field
\begin{eqnarray}
R_{ab,c,d}{}^a=0, \label{eq:3riccieq}
\end{eqnarray}
are
satisfied.
These equations can be derived in the same way as in
\cite{Hanada:2006gg}
\cite{Wess:1977fn}.

So far, we have analyzed the elimination of the unphysical degrees of 
freedom
by imposing constraints. We have found that the equation of motion for 
spin-3
field (\ref{eq:3riccieq}) is expressed in terms of the second order
derivatives of $e_{a,bc}$ and the one for spin-$\frac{5}{2}$ field
(\ref{eq:spin5/2eomc})
is expressed in terms of the first
order derivatives of $e_{a,b\alpha}$ respectively.
They are symmetric under permutations of the vector indices.
However, these constraints are not enough to
 eliminate all the unphysical degrees of
freedom.
The equations  (\ref{eq:spin5/2eom})
and (\ref{eq:3symtensoreom})
are expressed in terms of the totally symmetric tensor fields
$\phi_{abc}$ and $\psi_{ab,\alpha}$, but the equations
(\ref{eq:spin5/2eomc})
and (\ref{eq:3riccieq}) are expressed in terms of $e_{a,bc}$ and
$e_{a,b\alpha}$,
which have parts that are not totally symmetric.
Thus,
we should eliminate these degrees of freedom
in order to show that the lowest components of
the equations (\ref{eq:spin5/2eomc})
and (\ref{eq:3riccieq}) coincide with the equations
(\ref{eq:spin5/2eom}) and (\ref{eq:3symtensoreom}), respectively.
We will do these in the next subsubsection.


\subsubsection{Gauge fixing}

There are two kinds of gauge symmetries {\bf 1.} dynamical gauge symmetries
{\bf 2.} auxiliary gauge symmetries.
A dynamical gauge symmetry has an action on a dynamical gauge field
defined in (\ref{eq:dyn3}) and (\ref{eq:dyn52}), while an auxiliary gauge
symmetry
does not act on any of the dynamical gauge fields.
An auxiliary gauge symmetry generates shifts of auxiliary gauge fields
that are not determined in terms of the dynamical gauge fields by
solving the constraints.
These undetermined components are
exactly those which we have mentioned in the last part of the previous
subsubsection.
Thus, we should eliminate these degrees of freedom
by fixing gauge symmetries.
Recall that gauge symmetries are embedded in the superunitary symmetry of
the
supermatrix model (\ref{eq:mmunsym}).
We summarize the gauge transformations as follows :
\begin{enumerate}
\item[{\bf 1.}] Dynamical gauge transformations
\begin{itemize}
\item $\Lambda=\lambda^{ab}\partial_a\partial_b$ generates
\begin{eqnarray}
 \delta e_{a,bc}=\partial_a
\lambda_{bc}, \quad \delta ({ others})=0,
\label{eq:dyngt3}
\end{eqnarray}
where the parameter $\lambda_{ab}$ satisfies
$\lambda_{ab}=\lambda_{ba}$.
\item $\Lambda=\xi^{a,\alpha}\partial_a\nabla_\alpha$ generates
\begin{eqnarray}
 \delta e_{a,b,\alpha}=\partial_a\xi_{b,\alpha},\quad \delta (others)=0.
\label{eq:gaugetd}
\end{eqnarray}
\end{itemize}

\item[{\bf 2.}] Auxiliary gauge transformations
\begin{itemize}
\item
     $\Lambda=\tilde{\lambda}^{a,bc}(\partial_a\O_{bc}+\O_{bc}\partial_a)$
generates
\begin{eqnarray}
 \delta e_{a,bc}=\tilde{\lambda}_{b,ac}
+\tilde{\lambda}_{c,ab}, \quad \delta \omega_{a,b,cd}=
\partial_a\tilde{\lambda}_{b,cd},\label{eq:gaugetab1}
\quad \delta(others) =0,
\end{eqnarray}
where $\lambda_{a,bc}$ satisfies $\lambda_{a,bc}=-\lambda_{a,cb}$.
\item $\Lambda=\tilde{\lambda}^{ab,cd}(\O_{ab}\O_{cd}+\O_{cd}\O_{ab})$
generates
\begin{eqnarray}
 \delta \omega_{a,b,cd}=\tilde{\lambda}_{ab,cd}
+\tilde{\lambda}_{cd,ab},\quad
\delta \Omega_{a,bc,de}=\partial_{a} \tilde{\lambda}_{bc,de},
\quad \delta(others)=0,
\label{eq:gaugetab2}
\end{eqnarray}
where $\lambda_{ab,cd}$ satisfies 
$\lambda_{ab,cd}=-\lambda_{ba,cd}=-\lambda_{ab,dc}$.
\item $\Lambda=\tilde{\xi}^{ab,\alpha}(\O_{ab}\nabla_\alpha+
\nabla_\alpha\O_{ab})$ generates
\begin{eqnarray}
&& \delta e_{a,b,\alpha}=\tilde{\xi}_{ab,\alpha},\quad
\delta
\omega_{a,\alpha,cd}=\partial_a\tilde{\xi}_{cd,\alpha},
\quad \delta (others)=0,
\label{eq:gaugeta}
\end{eqnarray}
where
$\xi_{ab,\alpha}$ satisfies
$\xi_{ab,\alpha}=-\xi_{ba,\alpha}$.
\end{itemize}
\end{enumerate}

Under the dynamical gauge transformations (\ref{eq:dyngt3}) and
(\ref{eq:gaugetd}), the rank-3 totally symmetric tensor field
$\phi_{abc}(x)$ defined in (\ref{eq:dyn3})
and rank-2 totally symmetric
tensor-spinor field
$\psi_{ab,\alpha}(x)$ defined in (\ref{eq:dyn52}) transform as follows
:
\begin{eqnarray}
\delta \phi_{abc}(x)&=&\partial_a \lambda_{bc}(x)
+\partial_{b}\lambda_{ca}(x)
+\partial_c \lambda_{ab}(x),\\
\delta \psi_{ab,\alpha}(x) &=&
\partial_a \xi_{b,\alpha}(x)+\partial_b \xi_{a,\alpha}(x).
\end{eqnarray}
These correspond to (\ref{eq:symtengaugetdb}) and
(\ref{eq:symtengaugetd}),  respectively.
They are consistent with the identifications (\ref{eq:dyn3})  and
(\ref{eq:dyn52}).

As we will now show, using
the auxiliary  gauge transformations (\ref{eq:gaugetab1}),
(\ref{eq:gaugetab2}) and (\ref{eq:gaugeta}), we
can eliminate the parts of $e_{a,bc}$ and
$e_{a,b\alpha}$ that are not totally
symmetric in the vector indices,
and we can express dynamical variable in terms of $\phi_{abc}$ and
$\psi_{ab,\alpha}$.
We first fix the gauge symmetry (\ref{eq:gaugeta}).
Gauge fixing can be done by transforming
$e_{a,b\alpha}\rightarrow
\hat{e}_{a,b\alpha}=e_{a,b\alpha}+\tilde{\xi}_{ab,\alpha}$,
with choosing the parameter
$\tilde{\xi}_{ab,\alpha}$
as
\begin{eqnarray}
\tilde{\xi}_{ab,\alpha}=-e_{a,b,\alpha} +\psi_{ab,\alpha}
+\frac{1}{2}(\gamma_a)_{\alpha\beta}
\spsi_{b,}{}^\beta-\frac{1}{2}(\gamma_{b})_{\alpha\beta}\spsi_{
a,}{}^\beta -\frac{1}{3}\gamma_{ab}\psi'_\alpha
\end{eqnarray}
Carrying out this transformation, we can remove the part of
$e_{a,b\alpha}$
that is not
totally symmetric in the vector indices.
Substituting $\hat{e}_{a,b\alpha}$ into the equation
(\ref{eq:spin5/2eomc})
we can show that the equation
\begin{eqnarray}
(\gamma^a)_{\alpha\beta} C_{ab,c}{}^\beta=0
\end{eqnarray}
coincides with (\ref{eq:spin5/2eom}).
Next, we fix the gauge symmetries (\ref{eq:gaugetab1}) and
(\ref{eq:gaugetab2}).
Gauge fixing can be done by transforming $e_{a,bc}\rightarrow
\varepsilon_{a,bc}=e_{a,bc}+\tilde{\lambda}_{b,ac}+\tilde{\lambda}_{c,ab}$ 
and
$\omega_{a,b,cd}\rightarrow 
w_{a,b,cd}=\omega_{a,b,cd}+\tilde{\lambda}_{ab,cd}
+\tilde{\lambda}_{cd,ab}$, by  choosing the parameters 
$\tilde{\lambda}_{a,bc}$ and
$\tilde{\lambda}_{ab,cd}$ as we did in  \cite{Saitou:2006ca}.
In  \cite{Saitou:2006ca},
carrying out the gauge transformation by those parameters,
we have removed the part of $e_{a,bc}$
that is not totally symmetric in the vector
indices, and have shown that the equation
\begin{eqnarray}
R_{ab,c,d}{}^a=0
\end{eqnarray}
coincides with (\ref{eq:3symtensoreom}).
Thus, we have shown that with all these constraints and gauge fixing,
we obtain the free theory of the supermultiplet (3, 5/2).

\vspace{\baselineskip}

Before we close this section, we comment on the
generalization
of what we have done to a
massless supermultiplet with spins ($s, s-\frac{1}{2}$).
In order to deal with the  supermultiplet,
we keep track of the $(s-1)$th order polynomials of the operators
in $\nabla_A$ and $\O_{ab}$ :
\begin{eqnarray}
 \A_A = i\nabla_A
 &+& (i)^{s-1}e_{A,}{}^{A_1\cdots A_{s-1}}(z)\nabla_{A_1}\cdots
 \nabla_{A_{s-1}}
 \nonumber\\
  & +& \fr{(i)^{s-1}}{s-1}\omega_{A,}{}^{A_1\cdots A_{s-2},b_1
c_1}(z)
 \{
\nabla_{A_1}\cdots\nabla_{A_{s-2}}\O_{b_1c_1}
\}\nonumber\\
 & +& \fr{(i)^{s-1}}{(s-1)(s-2)}
 \Omega_{A,}{}^{A_1\cdots A_{s-3},b_1c_1,b_2c_2}(z)
\{\nabla_{A_1}\cdots\nabla_{A_{s-3}}
\O_{b_1c_1}\O_{b_2c_2}\}
 \nonumber\\
 &+&\fr{(i)^{s-1}}{(s-1)(s-2)(s-3)}
\tilde{\Omega}_{(1),A,}{}^{A_1\cdots A_{s-4},
b_1c_1,b_2 c_2,
 b_3 c_3}(z)\{\nabla_{A_1}\cdots\nabla_{A_{s-4}}
 \O_{b_1c_1}\O_{b_2c_2}\O_{b_3c_3}
 \} \nonumber\\
 && \quad\quad \vdots
 \nonumber\\
 &+& \fr{(i)^{s-1}}{(s-1)!}
 \tilde{\Omega}_{(s-3),  A}{}^{b_1c_1,\cdots,b_{s-1}c_{s-1}}(z)
 \{\O_{b_1c_1}\cdots\O_{b_{s-1}c_{s-1}}\}.
 \label{eq:spinsaexp}
 \end{eqnarray}
From the discussion in this section,
it seems that $\tilde{\Omega}_{(i)} (z)(i=1,\cdots,s-3)$
are not necessary to construct a superfield formulation
of the supermultiplet $(s,s-\frac{1}{2})$.
We set these auxiliary fields to zero :
$\tilde{\Omega}_1(z)=\cdots=\tilde{\Omega}_{s-3}(z)=0$.
Starting from this $\A_A$ we may construct a superfield formulation
of the massless supermultiplet $(s,s-\frac{1}{2})$
using the same method
as the one we have employed in this section.


\section{Supermatrix model}

Now, with the superfield formulation of the 
massless supermultiplet (3, 5/2), 
we can  compare the results with those of the supermatrix model.
Imposing the constraints in section 3, we obtain
the equations of motion of the supermatrix model
\begin{eqnarray}
{[} \A^a,[\A_a,\A_b {]}{]}
&=&{[} \partial^a,C_{ab,}{}^{c,\gamma}\partial_c
\nabla_\gamma+R_{ab}{}^{c,de}(\partial_c\O_{de}+\O_{de}\partial_c) {]}
\nonumber\\
&=&
( \partial^a
C_{ab}{}^{c\gamma})\partial_c\nabla_\gamma
+(R_{ab,}{}^{c,da})\partial_c\partial_d
+(\partial^a R_{ab,}{}^{c,de})(\partial_c\O_{de}+\O_{de}\partial_c)=0.
\label{eq:smmeom}
\end{eqnarray}
The equation $R_{ab,}{}^{c,da}=0$ coincides with (\ref{eq:3riccieq}).
$\partial^a R_{ab,}{}^{c,de}=0$  follows from
the superspace Bianchi identity (\ref{eq:4b13}) by contracting $a$ and $e$.
$\partial^a C_{ab,}{}^{c\gamma}=0$ is obtained by multiplying
$(\gamma^a)^{\gamma\alpha}\nabla_\gamma$ to (\ref{eq:4b22}).
Therefore, we have shown that solutions of the equations of motion for the
massless supermultiplet (3, 5/2) satisfy the equations of motion of the
supermatrix model (\ref{eq:smmeom}).


\section{Conclusions and future works}

In this paper, we have studied the relation between a supermatrix model and
 the free $4D$, $\N=1$ supersymmetric field theory of
a massless supermultiplet with spins (3, 5/2) on the basis of 
\cite{Hanada:2006gg}.
In order to do this,  we have constructed a superfield formulation of
the supermultiplet.
Then, we have shown that solutions of the equations of motion
for the supermultiplet  satisfy the equations of motion of the
supermatrix model.
It is difficult to show the converse that is 
to derive the equations of motion for the supermultiplet from 
the equations of motion of the supermatrix model.
We may generalize what we have done in this paper to
  the supermultiplet $(s,s-\frac{1}{2})$ using the same method
as the one we have employed.

There are several things which should be studied further.
One is to investigate the tensor fields which are not totally symmetric
in the spacetime vector indices can be included in  matrix models.
Viewed from matrix models, field strengths should be introduced as
independent degrees of freedom.
There is a possibility that ``field strengths'' propagate as tensor
fields
that are not totally symmetric.
This possibility has been studied in  \cite{Furuta:2006kk}.
The authors have investigated that the fields appear as the coefficients of 
terms
linear in the covariant derivative and local Lorentz generators.
They have found that some components of the torsion can be
identified with a scalar and rank-2 antisymmetric tensor field, and
have shown that the equations of these fields can be derived from that of
a matrix model.
It is interesting to extend this analysis to the fields that appear as
coefficients of higher order terms  in covariant derivative and
local Lorentz generators.

Another thing to be pursue is
to construct the interacting massless higher spin gauge field theory.
 Difficulties associated with the
requirement of gauge invariance can be overcome by using the
matrix model because it has gauge invariance manifestly.


\paragraph{Acknowledgment}

I am grateful to Y.~Baba, N.~Hatano, N.~Ishibashi, K.~Murakami,
Y.~Satoh and K.~Yamamoto for useful
discussions.


\appendix
\def\thesection{Appendix \Alph{section}}


\section{Summary of constraints}
\begin{itemize}
\item[{\bf 1.}] Constraints on component fields :
\begin{eqnarray}
&& e_{a,b}{}^b=0, \label{eq:compotl} \\
&& (\gamma^b)_\alpha{}^\beta e_{a,b,\beta}=0 \label{eq:compogtl}.
\end{eqnarray}

\item[{\bf 2.}] Constraints on superfield strengths

Off-shell constraints :
\begin{eqnarray}
&& C_{ab,}{}^{cd}=
C_{ab,}{}^{\gamma\delta}=0,
\qquad C_{a\alpha,}{}^{cd}=
C_{a\alpha,}{}^{\gamma\delta}= 0,
\label{eq:c23}\\
&& C_{\alpha\beta,}{}^{cd}= 2i(\gamma^a)_{\alpha\beta}e_{a,}{}^{cd}, \quad
C_{\alpha\beta,}{}^{c\gamma}=2i(\gamma^a)_{\alpha\beta}e_{a,}{}^{c\gamma}, 
\quad
C_{\alpha\beta,}{}^{\gamma\delta}=2i(\gamma^a)_{\alpha\beta}e_{a,}{}^{\gamma\delta}.
\label{eq:c33}
\end{eqnarray}
\begin{eqnarray}
R_{ab,}{}^{\gamma,cd}=
R_{a\alpha}{}^{\gamma,cd}=0,\qquad
R_{\alpha\beta}{}^{\gamma,cd}=2i(\gamma^a)_{\alpha\beta}
\omega_{a,}{}^{ \gamma,cd}.\label{eq:c34}
\end{eqnarray}
\begin{eqnarray}
F_{ab,}{}^{cd,ef}=
F_{a\alpha}{}^{cd,ef}=0,\qquad
F_{\alpha\beta}{}^{cd,ef}= 2i(\gamma^a)_{\alpha\beta}
\Omega_{a,}{}^{cd,ef}. \label{eq:c35}
\end{eqnarray}

On-shell constraints :
\begin{eqnarray}
C_{a\alpha}{}^{c\gamma}&=&0.\label{eq:eomc1}
\end{eqnarray}

\item[{\bf 3.}] Constraints on ``pure gauge'' field :
\begin{eqnarray}
\chi_{bc,de}=-\frac{1}{3}\omega_{[bc],de}, \label{eq:puregb}
\end{eqnarray}
\begin{eqnarray}
\eta_{cd,\gamma}=-e_{[c,d],\gamma}. \label{eq:puregf}
\end{eqnarray}

\end{itemize}


\section{Bianchi identities}
We give the
superspace Bianchi identities subject to the constraints
(\ref{eq:compotl}), (\ref{eq:compogtl}),
(\ref{eq:c23}), (\ref{eq:c33}), (\ref{eq:c34}),
(\ref{eq:c35}), (\ref{eq:eomc1}), (\ref{eq:puregb}) and (\ref{eq:puregf}).
\begin{itemize}
\item[{\bf 1.}] $[\A_a,[\A_b,\A_c]]+[\A_b,[\A_c,\A_a]]+[\A_c,[\A_a,\A_b]]=0$
   gives
\begin{eqnarray}
&& \partial_{[a}C_{bc]}{}^{d\delta}=0 ,\label{eq:4b11}\\
&& R_{[ab,}{}^{d,e}{}_{c]}=0, \label{eq:4b12}\\
&& \partial_{[a}R_{bc]}{}^{d,ef}=0. \label{eq:4b13}
\end{eqnarray}
\item[{\bf 2.}] $[\A_a,[\A_b,\A_\alpha]]+[\A_b,[\A_\alpha,\A_a]]
+[\A_\alpha,[\A_a,\A_b]]=0$ gives
\begin{eqnarray}
&&
2i(\gamma^d)_{\alpha\beta}C_{ab,}{}^{c\beta}+\frac{1}{2}R_{b\alpha,}{}^{c,d}{}_a
-\frac{1}{2}R_{a\alpha,}{}^{c,d}{}_b=0,\label{eq:4b21}\\
&& \nabla_\alpha C_{ab}{}^{c\beta}
+\frac{1}{2}R_{ab}{}^{c,de}(\gamma_{de})_\alpha{}^\beta=0,\label{eq:4b22}\\
&& \nabla_\alpha R_{ab}{}^{c,de}+\partial_a R_{b\alpha,}{}^{c,de}
-\partial_b R_{a\alpha,}{}^{c,de}=0.\label{eq:4b23}
\end{eqnarray}
\item[{\bf 3.}] $[\A_a,\{\A_\alpha,\A_\beta\}]+\{\A_\alpha,[\A_\beta,\A_a]\}
-\{\A_\beta,[\A_a,\A_\alpha]\}=0$ gives
\begin{eqnarray}
&& \tilde{R}_{\alpha\beta,}{}^{c,d}{}_e=0,\label{eq:4b31}
\\
&&
2(\gamma^b)_{\alpha\beta}C_{ab,}{}^{c\gamma}+\frac{1}{2}R_{a\beta,}{}^{c,de}
(\gamma_{de})_\alpha{}^{\gamma}+\frac{1}{2}R_{a\alpha,}{}^{c,de}
(\gamma_{de})_\beta{}^{\gamma}=0, \label{eq:4b32}\\
&& 2(\gamma^b)_{\alpha\beta}R_{ab}{}^{c,de}+
\nabla_\alpha R_{a\beta,}{}^{c,de}+\nabla_\beta R_{a\alpha,}{}^{c,de}=0.
\label{eq:4b33}
\end{eqnarray}
\item[{\bf 4.}] $[\A_\alpha,\{\A_\beta,\A_\gamma\}]+
[\A_\beta,\{\A_\gamma,\A_\alpha\}]+[\A_\gamma,\{\A_\alpha,\A_\beta\}]=0$
gives
\begin{eqnarray}
&&\tilde{R}_{(\alpha\beta}{}^{c,de}(\gamma_{de})_{\gamma)}{}^{\delta}=0,
\label{eq:4b41}
\\
&& 2(\gamma^a)_{(\alpha\beta}R_{a\gamma)}{}^{c,de}
+\nabla_{(\alpha}\tilde{R}_{\beta\gamma)}{}^{c,de}=0.
\label{eq:4b42}
\end{eqnarray}
\end{itemize}


\section{Solution of the Bianchi identities}

In this appendix, we give the results of an off-shell superfield
formulation for the theory of a   massless supermultiplet with spins (3, 
5/2).
To construct the formalism, we should impose the off-shell constraints
which reduce the number of components, and solve  Bianchi identities
subject to the off-shell constraints.
We can find that the Bianchi identities reduce the number of
independent superfields to one complex vector field $R_a$,
one real  symmetric-traceless tensor $G_{ab}$ and one chiral
superfield $W_{a,\alpha\beta\gamma}$.
We can find
explicit expressions for the superfield strengths
in terms of these superfields.

As in the case of supergravity, the lowest components
 of $R_a$ and $G_{ab}$ with respect to $\theta$ are physical degrees of
freedom.
The counting of field components are as follows :

{ \begin{tabular}{ccccccccc}
 {\bf Bosonic }  & $\phi_{abc}(x)$ &    $\lambda_{ab}(x)$  &
$R_{a}(x)$ &  $G_{ab}(x)$ &  \\
& +20 &   $-$9 & +8 & +9 & =28 \\
{\bf Fermionic}&  $\psi_{ab,\alpha}(x)$ & $\xi_{a,\alpha}(x)$ & \\
& +40 & $-$12 & & & =28
\end{tabular}}

Therefore, the number of bosonic and fermionic degrees of freedom
are equal.

\vspace{\baselineskip}

Here we use a two spinor notation of \cite{Wess:1992cp}.
The coordinates of flat superspace are denoted by
$z^A=(x^a,\theta^\alpha,\bar{\theta}^{\dot{\alpha}})$.
Latin indices $a$ denote Lorentz tensor indices, Greek indices
($\alpha,\dal$)
denote spinor indices.
Covariant derivatives in flat superspace are defined as follows
\begin{eqnarray}
\nabla_a&=&\partial_a, \\
\nabla_\alpha&=&\frac{\partial}{\partial\theta^\alpha}
+i\sigma^a{}_{\alpha\dal}
\bth^{\dal}\partial_a,
\\
\ovnab_{\dal}&=&-\frac{\partial}{\partial\bth^{\dal}}-i\theta^\alpha
\sigma^a{}_{\alpha\dal}\partial_a.
\end{eqnarray}

We list the results of the off-shell constraints and the solution of
superspace Bianchi
identities.

\vspace{\baselineskip}

{\bf Constraints}

We impose constraints on superfield strengths.
The commutators of $\A_A$ can be written as follows :
\begin{eqnarray}
{[}\A_A,\A_B{\}}= &&
-iC_{AB}{}^{CD}\nabla_C\nabla_D
-\frac{i}{2}
R_{AB}{}^{D,ef}(\nabla_D\O_{ef}
+\O_{ef}\nabla_D)\nonumber
\\ &&
-\frac{i}{2}F_{AB}{}^{cd,ef}(\O_{cd}\O_{ef}+\O_{ef}\O_{cd}).
\end{eqnarray}
We choose the following constraints on superfield strengths :
\begin{eqnarray}
&&C_{ab,cd}=C_{ab,\gamma\delta}=C_{ab,\gamma\dde}=0,\qquad
C_{a\alpha,cd}=C_{a\alpha,\gamma\delta}
=C_{a\alpha,\dga\dde}=C_{a\alpha,\gamma\dde}=0\\
&&C_{\alpha\beta,cd}=C_{\alpha\beta,d\delta}=C_{\alpha\beta,d\dde}
=C_{\alpha\beta,\dga\dde}=C_{\alpha\beta,\gamma\dde}
=C_{\alpha\beta,\gamma\delta}=0\\
&& C_{\alpha\dal,cd}=2i(\sigma^a)_{\alpha\dal}e_{a,cd},\quad
C_{\alpha\dal,d\delta}=2i(\sigma^a)_{\alpha\dal}e_{a,d\delta},\\
&&
C_{\alpha\dal,\gamma\delta}=2i(\sigma^a)_{\alpha\dal}e_{a,\gamma\delta},\quad
C_{\alpha\dal,\delta\dde}=2i(\sigma^a)_{\alpha\dal}e_{a,\delta\dde}.\\
&& R_{ab,\delta,cd}=0,\qquad R_{a\alpha,\delta,cd}=R_{a\alpha,\dde,cd}=0,
\qquad  R_{\alpha\beta,\delta,cd}=R_{\alpha\beta,\dde,cd}=0,
\\
&& R_{\alpha\dal,\delta,cd}=2i(\sigma^a)_{\alpha\dal}\omega_{a,\delta,cd},\\
&& F_{ab,cd,ef}=0, \qquad  F_{a\alpha,cd,ef}=0, \qquad
F_{\alpha\beta,cd,ef}=2i(\sigma^a)_{\alpha\dal}\Omega_{a,cd,ef},
\end{eqnarray}
\begin{eqnarray}
R_{\alpha\dal,\delta}{}^{\dde}{}_{,\beta\gamma}=0, \qquad
C_{\alpha\dal\beta,\delta}{}^{\dal}{}_\gamma=0, \label{eq:tracecnrtoff}
\end{eqnarray}
and their complex conjugates.
Here, we define $R_{\alpha\dal,\delta\dde,\beta\gamma}\equiv
(\sigma^d)_{\delta\dde}R_{\alpha\dal,d,\beta\gamma}$ and
$C_{\alpha\dal\beta,\delta\dde\gamma}\equiv
(\sigma^d)_{\delta\dde}C_{\alpha\dal\beta,d\gamma}$.

\vspace{\baselineskip}

{\bf Solution of the Bianchi identities}

\begin{itemize}
\item[{\bf 1.}] Constraints on the superfields $W$, $G$ and $R $
\begin{eqnarray}
&& \ovnab_{\dal}R_c=0, \label{eq:csf11}\\
&& \nabla^\alpha G_{c,\alpha\dbe}=\nabla_{\dbe}R^\dag_c ,
\label{eq:csf12}\\
&& \ovnab_{\dal}W_{c,\beta\gamma\delta}=0, \label{eq:csf13}\\
&& \nabla^{\alpha}W_{c,\alpha\beta\delta}
+\frac{i}{2}(\nabla_{\beta\dbe}G_{c,\delta}{}^{\dbe}
+\nabla_{\delta\dbe}G_{c,\beta}{}^{\dbe})=0,\label{eq:csf14}\\
&& G_{c,\alpha\dal}^\dag=G_{c,\alpha\dal},\label{eq:csf15}\\
&& W_{c,\alpha\beta\delta}^\dag = \overline{W}_{c,\dal\dbe\dde},
\label{eq:csf16}\\
&& G_{\gamma\dga,\alpha\dal}=(\sigma^c)_{\gamma\dga}G_{c,\alpha\dal}
=G_{(\gamma\alpha)(\dga\dal)},\label{eq:csf17}
\end{eqnarray}
and their complex conjugates.

As a consequence of the constraints (\ref{eq:tracecnrtoff}),
$G_{a,b}$ has only the traceless symmetric components.

\item[{\bf 2.}] Torsion
\begin{eqnarray}
&& C_{\alpha\dal\dbe,c\delta}= (\sigma^a)_{\alpha\dal}C_{a\dbe,c\delta}
=-2i\epsilon_{\dbe\dal}\epsilon_{\alpha\delta}R_c,\label{eq:torsf21}\\
&& C_{\alpha\dal\beta,c\delta}=(\sigma^a)_{\alpha\dal}C_{a\beta,c\delta}
=\frac{i}{4}(\epsilon_{\alpha\delta}G_{c,\beta\dal}-3\epsilon_{\beta\alpha}
G_{\delta\dal}-3\epsilon_{\beta\delta}G_{c,\alpha\dal}),\label{eq:torsf22}
\\ && C_{\alpha\dal\beta\dbe,c\dde}=
-2\epsilon_{\alpha\beta}\overline{W}_{\dal\dbe\dde,c}
 -\frac{1}{2}\epsilon_{\alpha\beta}
(\epsilon_{\dde\dbe}\nabla^\epsilon G_{c,\epsilon\dal}+\epsilon_{\dde\dal}
\nabla^\epsilon G_{c,\epsilon\dbe})
\nonumber\\
&& \hspace{80pt} +\frac{1}{2}\epsilon_{\dal\dbe}(\nabla_\alpha
G_{c,\beta\dde}
+\nabla_\beta G_{c,\alpha\dde}),\label{eq:torsf23}
\end{eqnarray}
and their complex conjugates.

\item[{\bf 3.}] Curvature
\begin{eqnarray}
&& R_{\dal\dbe,c,\dde\deps}=4(\epsilon_{\dal\deps}\epsilon_{\dbe\dde}
+\epsilon_{\dbe\deps}\epsilon_{\dal\dde})R_c,\label{eq:curvsf31}\\
&& R_{\dal\dbe,c,\delta\epsilon}=0,\label{eq:curvsf32}\\
&& R_{\alpha\dal,c,\beta\delta}=\epsilon_{\beta\alpha}G_{c,\delta\dal}
+\epsilon_{\delta\alpha}G_{c,\beta\dal},\label{eq:curvsf33}\\
&& R_{\alpha\dal\beta,c,\delta\epsilon}=
\frac{i}{2}(\epsilon_{\beta\alpha}\nabla_\delta+ \epsilon_{\beta\delta}
\nabla_\alpha)G_{c,\epsilon\dal}
+\frac{i}{2}(\epsilon_{\beta\delta}\nabla_\epsilon +
\epsilon_{\beta\epsilon}
\nabla_\alpha)G_{c,\delta\dal}\nonumber\\
&& \hspace{60pt}
+i(\epsilon_{\epsilon\beta}\epsilon_{\alpha\delta}+
\epsilon_{\delta\beta}\epsilon_{\alpha\epsilon})\nabla^\zeta G_{c,\zeta\dal}
,\label{eq:curvsf34}
\\
&&  R_{\alpha\dal\beta,c,\dde\deps}=  4i\epsilon_{\beta\alpha}
\overline{W}_{c,\dal\dde\deps}+\frac{i}{2}
(\epsilon_{\dal\dde}\nabla_\beta G_{c,\alpha\deps}+\epsilon_{\dal\deps}
\nabla_\beta G_{c,\alpha\dde}),\label{eq:curvsf35}
\\&&
R_{\alpha\dal\beta\dbe,c,\delta\dde\epsilon\deps}=
(\sigma^a)_{\alpha\dal}(\sigma^b)_{\beta\dbe}(\sigma^d)_{\delta\dde}
(\sigma^e)_{\epsilon\deps}R_{ab,c,de}
\\
&&
\hspace{60pt}
=4\epsilon_{\alpha\beta}\epsilon_{\delta\epsilon}
\overline{X}_{c,(\dal\dbe)(\dde\deps)}+4\epsilon_{\dal\dbe}
\epsilon_{\dde\deps}X_{c,(\alpha\beta)(\delta\epsilon)}
\nonumber\\
&& \hspace{70pt}
-4\epsilon_{\alpha\beta}\epsilon_{\dde\deps}
\overline{\Psi}_{c,(\dal\dbe)(\delta\epsilon)}
-4\epsilon_{\dal\dbe}\epsilon_{\delta\epsilon}
\Psi_{c,(\alpha\beta)(\dde\deps)}, \label{eq:curv36}\\
&&
\Psi_{c,(\alpha\beta)(\dde\deps)}=
\overline{\Psi}_{c,(\dde\deps)(\alpha\beta)}
,\label{eq:curv37}
\\ &&
X_{c,(\alpha\beta)(\delta\epsilon)}=
-\frac{1}{4}(\nabla_\alpha W_{c,\beta\delta\epsilon}+\nabla_\beta
W_{c,\delta\epsilon\alpha}+\nabla_\delta
W_{c,\epsilon\alpha\beta}+\nabla_\epsilon
W_{c,\alpha\beta\delta})\nonumber\\
&& \hspace{70pt}
+
\frac{1}{16}(\ovnab_{\dot{\zeta}}\ovnab^{\dot{\zeta}}R_c^\dag
+\nabla^{\zeta}
\nabla_{\zeta}R_c),\label{curv38}\\
&& \Psi_{c,(\alpha\beta)(\dde\deps)}=
\frac{i}{8}(\nabla_{\beta\dde}G_{c,\alpha\deps}+\nabla_{\alpha\dde}
G_{c,\beta\deps}+\nabla_{\beta\deps}G_{c,\alpha\dde}+
\nabla_{\alpha\deps}G_{c,\beta\dde})
\nonumber\\
&&
\hspace{70pt}
+\frac{1}{8}(\ovnab_{\dde}\nabla_\beta G_{c,\alpha\deps}
+\ovnab_{\dde}\nabla_\alpha G_{c,\beta\deps}
+\ovnab_{\deps} \nabla_\beta G_{c,\alpha\dde}
+\ovnab_{\deps}\nabla_\alpha G_{c,\beta\dde}),\label{eq:curvsf39}
\end{eqnarray}
and their complex conjugates.
\end{itemize}
All other components vanish.


\end{document}